\documentstyle[onecolumn]{mn} 
\begin{document}
\newcommand{\beq}{\begin{equation}}
\newcommand{\eeq}{\end{equation}}
\newcommand{\mg}{\marginpar} 
\newcommand{\ti}{\tiny} 
\newcommand{\ul}{\underline}
\newcommand{\nid}{\noindent}


\title[Fast and Accurate Computation...]
{Fast and Accurate Computation Tools\protect\\ 
for Gravitational Waveforms\protect\\
from Binary Stars with any Orbital Eccentricity}


\author[V. Pierro et al.]
{
V. Pierro $^{1}$, 
I.M. Pinto $^{1}$,
A.D. Spallicci $^{1}$,
E. Laserra $^2$, 
F. Recano $^2$.\\ 
$^1$Waves Group, Univ. of Sannio at Benevento, Italy\\
$^2$D.I.I.M.A., Univ. of Salerno, Italy
}

\maketitle


\begin{abstract}
The relevance of orbital eccentricity in the 
detection of gravitational radiation from 
(steady-state) binary stars is emphasized.
Computationally effective (fast and accurate) tools 
for constructing gravitational wave templates from binary
stars with any orbital eccentricity are introduced,
including tight estimation criteria of the pertinent truncation 
and approximation errors.
\end{abstract}

\section*{1 INTRODUCTION}  

Gravitational wave detection experiments in space, including
satellite Doppler-Tracking (Bertotti and Iess, 1999) and LISA (http://lisa.jpl.nasa.gov), 
will hopefully open a window on the low-frequency part
of the gravitational wave (henceforth GW) spectrum of cosmic origin. 
In these frequency bands, binary stars are among
the most promising continuous detectable source. 

A substantial fraction of binaries are expected to have orbits with 
{\em non negligible eccentricity} (Barone et al., 1988; Hils et al., 1992; 
Pierro and Pinto, 1996c) resulting into the emission of 
{\em several harmonics} of the fundamental orbital frequency.
The importance of this fact from the standpoint of signal detection
and estimation has been already noted.

For {\em coalescing binaries}, Pierro and Pinto (1996b) and 
Martel and Poisson (1999) pointed out
that neglecting residual
(albeit very small) orbital eccentricities may seriously
deteriorate  matched-filter detection performance.
Their results, obtained in the frame of the
simplest (newtonian) Peters Mathews (henceforth PM)  model 
(Peters and Mathews, 1963; Peters 1964;
Pierro and Pinto 1996c),
support the {\em qualitative} conclusion that residual
orbital eccentricities cannot be {\em bona fide} 
disregarded in  building  templates
for matched-filter detection
of gravitational wave chirps 
from inspiraling binaries\footnote{
The effect of a residual (tiny) orbital eccentricity on the
radiation emitted from an {\em inspiraling} binary system was also 
considered in (Moreno et al., 1994), with special emphasis on the possible
relevance of periastron advance. The results in (Moreno et al., 1994)
are unfortunately affected by several errors and misprints.}.

For {\em steady-state} binaries with non-zero
orbital eccentricity, on the other hand, 
using  circular-orbit waveform
templates, i.e. neglecting higher
order harmonics, implies a {\em potentially
large} loss of signal-to-noise ratio (henceforth SNR),
leading to significantly worse 
detector's performance, as will be 
shown in the sequel.

The main goals of the present paper are:

\begin{itemize}
\item[i)]{ to provide some quantitative hint for
validating the applicability of the simple  PM
model to steady-state binaries;}
\item[ii)]{ to gauge the loss in SNR
due to the simple circular-orbit assumption and, more
generally, to set some criteria for spectral waveform truncation;}
\item[iii)]{to introduce {\it efficient} 
(accurate and fast) computational tools for 
constructing gravitational waveform templates
for (steady-state) binary sources with {\em any} orbital
eccentricity.} 
\end{itemize} 

The paper is  organized as follows.
In {\em Sect. 2} we introduce some (dimensionless) 
parameters whereby the applicability 
of the PM model  to specific sources can be assessed.
In {\em Sect.s 3a} and {\em 3b} we review the GW
spectra and waveforms in the frame of the PM model. 
In {\em Sect.s 4a} and {\em 4b} we show how to evaluate the total
harmonic distortion due to spectral waveform truncation,
and introduce a modified Carlini Meissel expansion tool
for fast and accurate GW harmonics computation.
The results in this section can be readily extended,
in principle, to higher-order post-newtonian (henceforth PN) models.
As an application, in {\em Sect. 5} we apply our formalism 
to some paradigm eccentrical binary sources. 
Conclusions follow under {\em Sect. 6}. 
Technical developments are collected in {\em Appendix A} to {\em C}.

\section*{2 STEADY-STATE BINARIES: THE PETERS-MATHEWS MODEL}

The PM  model
for gravitational wave emission from binary 
systems in a Keplerian orbit was introduced in the sixties
(Peters and Mathews, 1963, Peters 1964), and recently
re-examined  (Pierro and Pinto, 1996a)
. It relies
on the following main assumptions:
i) point mass, ii) weak field, iii) slow motion, and iv) adiabatic evolution
(negligible change of the orbital parameters over each orbit).
These conditions can be checked in terms of
the following inequalities (Pierro and Pinto, 1996a):

\beq
\xi_1:=\frac{source~gravitational~radius}{aphastral~separation}=
2\chi^{-2/3}(1-e)^{-1} \ll 1,
\eeq
  
\beq
\xi_2:=\frac{aphastral~velocity}{velocity~of~light}=
\chi^{-1/3}
\left(
\frac{1+e}{1-e}
\right)^{1/2}
\ll 1,
\eeq
  
\beq
\xi_3:=sup
\left\{
\left|\frac{dT}{dt}\right|, 
\left|e^{-1}\frac{de}{dt}~T\right| 
\right \}
=\frac{152\pi}{15}(1-\Delta^2)\chi^{-5/3}\cdot
\left(1+\frac{121}{304}e^2\right)(1-e^2)^{-5/2}
\ll 1,
\eeq
where as already stated
$\chi = cT/\pi r_g$, $T$ being the orbital period,
$r_g=2G(M_1+M_2)/c^2$  is the  source gravitational radius,  
$M_{1,2}$ are  the companion masses,
$\Delta~=~|M_1-M_2|/(M_1+M_2)$ and $e$ is the eccentricity.

Tidal effects could be  neglected 
provided neither companion star fills its Roche lobe. 
Following Eggleton (1983), this translates into:

\beq
\xi_1 \ll \Lambda^{-1}
\frac{2}{1+\Delta}
\left\{
0.6+\left(\frac{1-\Delta}{1+\Delta} \right)^{2/3}
\ln \left [
1+\left(\frac{1-\Delta}{1+\Delta} \right)^{-1/3}
\right ]
\right\}^{-1},
\eeq
where $\Lambda$ is the ratio between the physical and
gravitational companion radius\footnote{Typical values of $\Lambda$
range from $10^4$ for white dwarfs down to $3$ for hadronic stars.
Further departures from the standard model
are expected due to the possible occurrence of
mass-transfer phenomena, which would be present in
closely-orbiting classical stars, as well as in
binaries where one companion is an accreting
collapsed object.}.

For most  steady-state binary systems, i.e. long before coalescence,
$\xi_1$ to $\xi_3$ above are fairly small (see e.g. {\em Sect. 5}),
and the PM model turns out to be perfectly adequate.

\section*{3a STEADY-STATE BINARIES: SPECTRA}

According to the PM model, the 
GW   power $\overline{\cal L}_{n}^{+,\times}$
radiated  at the $n^{th}$ harmonic
of the orbital frequency  by a
steady binary source can be 
conveniently cast into the following universal
form (Barone et al., 1988):

\beq
\overline{\cal L}_{n}^{+,\times}=
\frac{2G}{5c^5}\chi^{-10/3}
(1-\Delta^2)^2
G_{max}(e)
\overline{g}^{+,\times}(n,e)
\label{eq:lumen}
\eeq
where the superscripts $+,\times$ refer to the fundamental 
GW polarization states.
The  spectral power distribution is embodied in the
{\em universal} dimensionless functions
$\overline{g}^{+,\times}(n,e)$  shown in {\em fig.s 1.1-1.10} for $e=0(0.1)0.9$.
For circular orbits ($e=0$)  only the second harmonic
is emitted. 
The function $G_{max}(e)$ plotted in {\em fig. 2}
is the ratio between the total luminosity 
(sum over both polarizations) 
of the brightest GW spectral line,
and the total luminosity of a circular-orbit binary
having the same $\chi$ and $\Delta$.
The brightest spectral line is
the $N_{max}$-th harmonic of the orbital frequency,
where $N_{max}$ is a function of $e$ only, displayed in {\em  fig. 3}.

It is seen that for non circular orbits, {\em several}
spectral lines with {\em comparable} intensities are emitted.
Thus, use of the circular orbit waveform templates implies
a potentially sizeable loss in  the available signal power
and hence in  the SNR, which can spoil
the detector's performance.

\subsection*{3b WAVEFORMS}

The far-field metric deviation (TT gauge) in  the PM model is\footnote{
The  GW  field  can  also be obtained
by inverting  the Keplerian integral of motion
relating time to the true anomaly, and exploiting the
simple dependance of the radiated waveforms on this latter
(Wahlquist, 1987).
The referred procedure is purely numerical and,
to the best of our knowledge, its generalization
to higher PN order models is not immediate.}:

\beq
h_{\times} = \frac{\cos \vartheta}{\sqrt 2}
\left [ 2h_{xy} \cos 2\varphi - (h_{xx} - h_{yy}) \sin 2\varphi
\right ],
\label{eq:hx}
\eeq
\beq
h_{+} = \frac {1}{\sqrt 2}
\left\{
\frac{3+ \cos 2\vartheta}{4}
\left[
2h_{xy} \sin 2\varphi + (h_{xx}-h_{yy}) \cos 2\varphi
\right] -
\frac{1\!-\!\cos 2\vartheta}{4} 
(h_{xx} + h_{yy})
\right\},
\label{eq:hp}
\eeq
where the coordinates $\vartheta$ and $\varphi$
specify the direction of the observer in a
spherical polar system where the orbit lies
in the equatorial plane and the binary center of mass is at the origin.\\
The metric components in (\ref{eq:hx}), 
(\ref{eq:hp}) can be  expanded into Fourier series under the
{\em adiabatic assumption} that the orbital parameters
{\em do not} change appreciably over each orbit. Hence\footnote{
The unknown irrelevant phase at $t=0$ has been set to zero.}:

\beq
h_{xy} = \sum_{n=1}^{\infty} h_{xy}^{(n)} \sin \left( n \frac{2\pi}{T} t \right) ,
\label{eq:unos}
\eeq

\beq
h_{x \pm y} = \sum _{n=1}^{\infty} h_{x\pm y}^{(n)} \cos \left( n \frac{2\pi}{T} t \right),
\label{eq:dos}
\eeq
where  $h_{x \pm y}$ is a shorthand for $h_{xx} \pm h_{yy}$
(see {\em Appendix A}), 
  
\beq
h^{(n)}_{xy} =
h_{0}n(1-e^{2})^{1/2}
\left[
J_{n-2}(ne) + J_{n+2}(ne)
- 2J_{n}(ne)
\right],
\label{eq:prima}
\eeq

\beq
h^{(n)}_{x-y} = 2 h_{0} n
\left\{ 
J_{n-2}(ne) - J_{n+2}(ne)~-~2e
\left [J_{n-1}(ne) - J_{n+1}(ne) \right ] +
(2/n) J_{n}(ne)
\right \},
\label{eq:seconda}
\eeq

\beq
h^{(n)}_{x+y} = -4h_0
J_n(ne),
\label{eq:terza}
\eeq
and\footnote{Note that for $n=1$ eq.s (\ref{eq:prima}) and (\ref{eq:seconda}) contain
Bessel functions of order $-1$, for which $J_{-1}(x)=-J_1(x)$.}

\beq
h_{0} = \frac{cT}{4\pi r}
\frac{1-\Delta^{2}}{\chi^{5/3}}.
\label{eq:h0}
\eeq
For circular orbits one has simply:

\beq
h_{x-y}^{(n)}=2h_{xy}^{(n)}=4h_0 \delta_{n2},~~~
h_{x+y}^{(n)}=0,
\eeq
where $\delta_{pq}$ is the Kronecker symbol.

For {\em steady state} binaries 
the (Robertson) periastron advance\footnote{The relativistic
periastron advance was heuristically (i.e., inconsistenlty, from
the post-newtonian expansion view point)  included in
(Moreno et al., 1995).} does not produce sensible effects 
on the waveforms, and is thus deliberately ignored.
Inclusion of the periastron advance amounts to splitting
each GW spectral line  into a doublet 
at $\sim (2\pi/T) (1 \pm 6\chi^{-2/3})$, which {\em cannot}
be resolved unless  the signal is  Fourier-transformed 
over a timespan  $\sim \chi^{2/3} T~\mbox{sec}$.
This time is, e.g., $\sim 5\cdot 10^5$ years  and $\sim 2.8\cdot 10^5$ years
for $PSR1534+12$ and $PSR1913+16$, respectively.

\section*{4a SPECTRAL TRUNCATION AND APPROXIMATION ERROR}

In order  to discuss the effect of spectral truncation of
(\ref{eq:unos}) and (\ref{eq:dos}) on the available SNR
it is convenient to introduce
the total harmonic distortion (henceforth THD):
\beq
THD=
\frac
{
\|h-\stackrel{\sim}{h}\|
}{
\|h\|
}
=
\left(
\frac
{
\displaystyle{\sum_{n=1}^{\infty}}~
\left(
\stackrel{~}{h}^{(n)}-\stackrel{\sim}{h}^{(n)}
\right)^2
}{
\displaystyle{\sum_{n=1}^{\infty}}~
|h^{(n)}|^2
}
\right)
^{1/2},
\eeq
where $h$, $\stackrel{\sim}{h}$ represent the {\em exact} and 
{\em approximate} values of the metric tensor,
$h^{(n)}$, $\stackrel{\sim}{h}^{(n)}$
are the  Fourier coefficients of $h$, $\stackrel{\sim}{h}$, respectively, and
the $L^2$-norms are computed by taking the time average 
over one orbital period of the square of the argument, within the spirit of 
the adiabatic approximation.
If only $N_T$ harmonics are included, then

\beq
\stackrel{\sim}{h}^{(n)} =
\left\{
\begin{array}{l}
h^{(n)},~~~n \leq N_T,\\
0,~~~~~~~~~n > N_T
\end{array}
\right.,~~~~
THD=
\left(
1-
\frac{
\displaystyle{\sum_{n=1}^{N_T} \left| h^{(n)} \right|^2 }
}{
\displaystyle{\sum_{n=1}^{\infty}  \left| h^{(n)} \right|^2}
}
\right)^{1/2}.
\label{eq:D1}
\eeq 
It is readily recognized that $THD^2$  represents the fraction
of signal power which is {\em lost} as an effect of truncation\footnote{
The THD is closely related to the fitting factor FF
(Apostolatos, 1996) between the exact and spectral-truncated
(template) waveform. From the very definitions
one gets:
$$
FF~\sim~
1-\frac{THD~^2}{2}~+~{\cal O}(~THD~^3~).
$$.}.

In the most general case, where besides spectral truncation,
the Fourier coefficients are computed in {\em approximate} form
(as e.g.  in the next subsection), one has:
\beq
THD=
\left(
1
-
\frac
{
2~\displaystyle{\sum_{n=1}^{NT}}~h^{(n)}~\stackrel{\sim}{h}^{(n)}-
\displaystyle{\sum_{n=1}^{NT}}~|\stackrel{\sim}{h}^{(n)}|^2
}{
\displaystyle{\sum_{n=1}^{\infty}}~|h^{(n)}|^2 
}
\right)^{1/2},
\label{eq:D2}
\eeq
The harmonic distortions  $\mbox{THD}_{x \pm y}$, $\mbox{THD}_{xy}$ 
due to the spectral truncation of (\ref{eq:unos}), (\ref{eq:dos}) 
can be computed for any given $N_T$ using  Kapteyn's theory 
(Watson, 1966, ch. 17)  to evaluate in closed form
the infinite sums in (\ref{eq:D2}).
After some lengthy but simple algebra, one obtains (see Appendix B):
\beq
||h_{x+y}||^2~=~
\displaystyle{\sum_{n=1}^{\infty}}~
\left|~h^{(n)}_{x+y}~\right|^{~2}
~=~
8~\left[~(1-e^2)^{-1/2}~-1~\right],
\eeq
\beq
||h_{x-y}||^2~=~
\displaystyle{\sum_{n=1}^{\infty}}~
\left|~h^{(n)}_{x-y}~\right|^{~2}
~=~e^{-4}
\left\{
4~(1-e^2)^{-1/2}~(8-12~e^2+9~e^4)-
8~(e^2-2)^2
\right\},
\eeq
\beq
||h_{xy}||^2~=~
\displaystyle{\sum_{n=1}^{\infty}}~
~\left|~h^{(n)}_{xy}~\right|^{~2}
~=~
e^{-2}~(1-e^2)^{-1/2}~
\left\{~12+e^2~+~8~e^{-2}~\left[(1-e^2)^{3/2}-1\right]\right\},
\eeq

The corresponding harmonic distortions for the $TT$  metric components $h_{+}$,
$h_{\times}$  can be conveniently written 
as follows:
\beq
THD_{\times}=
\left\{
\left[
4 THD^{2}_{xy} \|h_{xy}\|^2 \cos^{2} 2\varphi+
THD^{2}_{x-y} \|h_{x-y}\|^2 \sin^{2} 2\varphi 
\right] \cdot \left[
4 \|h_{xy}\|^2 \cos^{2} 2\varphi+
\|h_{x-y}\|^2 \sin^{2}2\varphi
\right]^{-1}
\right\}^{1/2},
\label{eq:THDc}
\eeq
and:

\[
THD_{+}=
\left\{
\left[
\left(
\stackrel{}{3}\!+\!\cos 2\vartheta
\right)^2
\left(
   4THD^{2}_{xy} \|h_{xy}\|^2 \sin^{2} 2\varphi\!+\!
   THD^{2}_{x-y} \|h_{x-y}\|^2 \cos^{2} 2\varphi 
\right)
\!+\!
\left(
1\!-\!\cos 2\vartheta 
\right)^2
  THD^{2}_{x+y} \|h_{x+y}\|^2~+
\right.\right.
\]
\[
\left.\left.
+~
2~\!(1\!-\!\cos 2\vartheta)(3\!+\!\cos 2\vartheta)\cos 2\varphi
\left< 
(h_{x-y}-\stackrel{\sim}{h}_{x-y}),
(h_{x+y}-\stackrel{\sim}{h}_{x+y})
\right>
\right] \cdot \left[
\left(
\stackrel{}{3}\!+\!\cos 2\vartheta
\right)^2
\left(
   4 \|h_{xy}\|^2 \sin^{2} 2\varphi\!+\!
       \|h_{x-y}\|^2 \cos^{2} 2\varphi 
\right)\!+\!
\right.\right.
\]
\beq
\!+\!
\left.\left.
(1\!-\!\cos 2\vartheta)^2
  \|h_{x+y}\|^2\!+\! 
2~\!(1\!-\!\cos 2\vartheta)(3\!+\!\cos 2\vartheta)\cos 2\varphi
\left<\stackrel{}{h_{x-y}}, \stackrel{}{h_{x+y}}
\right>
\right]^{-1}
\right\}^{1/2},
\label{eq:THDp}
\eeq
where $\langle \cdot, \cdot \rangle$  is the scalar product
in $L^2_{[0,T]}$.  In order to evaluate  (\ref{eq:THDp})
the further infinite sum:

\beq
\left<
h_{x-y},h_{x+y}
\right> ~=~
\displaystyle{\sum_{n=1}^{\infty}}~
h^{(n)}_{x-y}~
h^{(n)}_{x+y}
~=~
-8~(1-e^2)^{-1/2}~
\left\{
~1+~(1-2~e^{-2})~
\left[1-~(1-e^2)^{1/2}\right]
\right\}.
\eeq
is needed, which is also readily obtained as explained
in Appendix B.

The harmonic distortions (\ref{eq:THDc}) and (\ref{eq:THDp})
can be sensible even at very low eccentricities ($e \leq .1)$.
Expanding (\ref{eq:THDc}) and (\ref{eq:THDp}) to lowest order in $e$ yields:
\beq
THD_\times=\frac{3~\sqrt{10}}{4}~e~+{\cal O}(e^3),~~~
\label{eq:THDc1}
\eeq
\beq
THD_+=
\frac{\sqrt{4(1-\cos 2\varphi)^2+
12(1-\cos 2\varphi)(3+\cos 2\varphi)\cos2\vartheta+
90(3+\cos 2\varphi)^2}
}{4(3+\cos 2\varphi)}~e~+ {\cal O}(e^3).
\label{eq:THDp1}
\eeq
The above simple expressions are fairly accurate for $e \leq .1$,
as seen, e.g., from {\em fig. 4}, where the angular averages of 
the approximate and exact 
harmonic distortion are drawn, and seen to be almost
indistinguishable and non-negligible.
The $(\vartheta,\varphi)-$dependent factor in
(\ref{eq:THDp1})  is plotted in {\em fig. 5}. Its 
average value over the sphere is exactly equal to the 
$(\vartheta,\varphi)-$independent factor in
(\ref{eq:THDc1}).

The obvious question is how many terms 
should be included in (\ref{eq:unos}) and  (\ref{eq:dos})
so as to keep both $THD_{+}$ and $THD_{\times}$ 
below some specified level,  for any $(\vartheta, \varphi)$.

To answer this question one may resort to the following
inequalities:
\beq
\max_{(\vartheta,\varphi)}THD_{\times} \leq max(THD_{xy}, THD_{x-y}),~~~
\max_{(\vartheta,\varphi)}THD_{+} \leq max(THD_{xy}, THD_{x-y}, THD_{x+y})~
\max_{(\vartheta,\varphi)}[Q(\vartheta,\varphi,e)].
\label{eq:ineq}
\eeq
where:
\[
Q(\vartheta,\varphi,e)\!=\!
\left\{
\left[
\left(
\stackrel{}{3}\!+\!\cos 2\vartheta
\right)^2
\left(
   4 \|h_{xy}\|^2 \sin^{2} 2\varphi\!+\!
       \|h_{x-y}\|^2 \cos^{2} 2\varphi 
\right)
\!+\!
\left(
1\!-\!\cos 2\vartheta 
\right)^2
       \|h_{x+y}\|^2~+
\right.\right.
\]
\[
\left.\left.
+~
2~\! | (1\!-\!\cos 2\vartheta)(3\!+\!\cos 2\vartheta)\cos 2\varphi |~
\|h_{x-y}\|  \|h_{x+y}\| \frac{}{}
\right] \cdot \left[
\left(
\stackrel{}{3}\!+\!\cos 2\vartheta
\right)^2
\left(
   4 \|h_{xy}\|^2 \sin^{2} 2\varphi\!+\!
       \|h_{x-y}\|^2 \cos^{2} 2\varphi 
\right)\!+\!
\right.\right.
\]
\beq
\!+\!
\left.\left.
(1\!-\!\cos 2\vartheta)^2
  THD^{2}_{x+y} \|h_{x+y}\|^2\!+\! 
2~\!(1\!-\!\cos 2\vartheta)(3\!+\!\cos 2\vartheta)\cos 2\varphi
\left<
\stackrel{}{h_{x-y}}, \stackrel{}{h_{x+y}}
\right>
\right]^{-1}
\right\}^{1/2},
\label{eq:Q}
\eeq
The first of (\ref{eq:ineq}) follows immediately from (\ref{eq:THDc}); 
the second one is obtained from (\ref{eq:THDp}) using Schwartz inequality.  
The supremum of the function $Q(\vartheta,\varphi,e)$
occurs   at $\vartheta=\pi/2$, $\varphi=m\pi$,  $\forall~e$,
where $Q(\pi/2,~m\pi,~e) \stackrel{<}{\sim}1.5$  (see {\em fig. 6}).

The truncation orders required to keep  $THD_{\times,+} \leq 0.01$, 
deduced from (\ref{eq:ineq})  are collected in {\em Table-I}.

\begin{center}
\begin{tabular}{||c|c||} \hline
$e_0$   &   $N_T$   \\ \hline
.1      &       4 \\
.2      &       6 \\
.3      &       8 \\
.4      &       11 \\
.5      &       15 \\
.6      &       22 \\
.7      &       36 \\
.8      &       68 \\
.9      &       206 \\ \hline\\
\end{tabular}
\\
{\em Table I - Truncation orders needed to keep $THD_{+,\times} \leq .1$.}
\end{center}

\section*{4a A GENERALIZED CARLINI-MEISSEL FORMULA}

A key issue for an efficient computation of
waveform-templates based on 
(\ref{eq:prima}), (\ref{eq:seconda}) and (\ref{eq:terza})
involves clever evaluation of terms like:

\beq
J_n(ne), J_{n\pm 1}(ne), J_{n\pm 2}(ne).
\eeq
It is well known that, in general, whenever the
argument and the order are close (here, in fact they are
proportional through the orbital eccentricity $e$), numerical
computation of Bessel functions either by series summation 
(Abramowitz and Stegun, 1968, ch. X), or by (re-normalized,
downward) recurrence (Press et al, 1992, Sect. 6.5)  
is  inefficient.
As a convenient alternative,
we suggest the following generalization 
of the well-known (see Watson, 1976, ch. XVII)
Carlini-Meissel (henceforth CM)  expansion:

\beq
J_{n \pm k}(ne) \sim J_{n}^{(CM)}(ne)\Psi_{\pm k}(n,e),
\eeq
where (see Appendix C for the detailed deduction): 

\beq
J_{n}^{CM}(ne) = \displaystyle{
\frac
{\displaystyle{\left(\frac{ne}{2}\right)^{n}}}
{n!}
} 
\left ( \frac{1+\sqrt{1-e^{2}}}{2}\right )^{-n} (1-e^{2})^{-1/4}
\cdot
\exp \left \{n\left [\sqrt{1 -e^{2}} -1 \right ] +
n^{-1}\left [\frac{-3e^{2} -2}{24(1-e^{2})^{3/2}} + \frac{1}{12}\right ]
\right \},
\label{eq:CM1}
\eeq

\beq
\Psi_{\pm k}(n,e) = \frac {n!}{(n\pm k)!}\left (\frac{ne}{1+\sqrt{1-e^{2}}}
\right )^{\pm k}~\cdot
exp \left \{\frac{1}{n}\left [ \frac{\mp k}{2}\frac{e^2}{1-e^2} +
\frac{k^{2}}{2} 
\left (1-\frac{1}{\sqrt{1-e^{2}}}\right )
\right ]
\right \}.
\label{eq:CM2}
\eeq

Using (\ref{eq:CM2}) to evaluate the Fourier coefficients
$\stackrel{\sim}{h}^{(n)}$  does  {\em not}  significantly
spoil the accuracy of the  waveforms. 
Indeed, spectral truncation according to Table-I still yields
THD values below $0.1$.

\section*{5 PROTOTYPE  SOURCES} 

As an application of the above, we wrote a code for
waveform template construction, and used it to compute
the waveforms for several prototype sources.
Taylor et al. (1993) provide data for 24 binary pulsars. 
In {\em Table II} we quote {\it PSR 1913+16} and {\it PSR 1534+12}, as 
possible paradigm sources for space detectors, being respectively
the most popular and closest known binary pulsars.
\begin{center}
\begin{tabular}{||c|c|c||} \hline
{\bf Binary}                            & {\bf 1534+12}          & {\bf 1913+16}     \\ \hline
Right ascension $B1950$                 & 15:34:47.686       &19:13:12.46769\\ \hline
Declination $B1950$                     & +12:05:45.23        &+16:01:08.0323 \\ \hline
Orbital inclination $i$ [degrees]       & 74                 & 45             \\ \hline
Distance [kpc]                          & 0.68               & 7.13         \\ \hline
Projected semimajor axis $a_{i}\sin i [light \cdot s]$&3.729468&2.3417592   \\ \hline
Eccentricity $e$                        & 0.2736779          & 0.6171308    \\ \hline
Orbital period $P_{b}$ [d]              & 0.4207372998       &0.322997462736\\ \hline
Companion masses $[M_{\odot}]$          & 1.34, 1.34         &1.42, 1.41    \\ \hline
$\xi _{1}$ ($10^{-6}$)                  & 3.4849             & 4.3102       \\ \hline
$\xi _{2}$ ($10^{-3}$)                  & 1.3200             & 1.4680       \\ \hline
$\xi _{3}$ ($10^{-14}$)                 & 7.6549             & 13.023       \\ \hline
$\Delta$                                & 0                  &$3.5336\times 10^{-3}$ \\ \hline
$\chi$                             &434777882.4767      &316085232.7313 \\ \hline
$h_{0}$ ($10^{-23}$)                    &16.518              &2.0575 \\ \hline\\
\end{tabular}
\\
{\em Table II - Paradigm compact binary sources}
\end{center}
The gravitational waveforms at $\vartheta=\varphi=0,45,90~deg$, 
computed using $8$ harmonics
for PSR1534+12 and $22$ harmonics for PSR1913+16
(consistent with {\em Table-I})
are displayed in {\em fig.s 7.1-7.15}, and 
{\em fig.s 8.1-8.15}, respectively. By comparison, the 
waveforms corresponding to $e=0$ are also drawn.

\section*{6 CONCLUSIONS}

The main results in this paper can be summarized as follows. 
Orbital eccentricity should not be neglected in detecting
gravitational waves from steady-state binaries, for which
the simple Peters Mathews model has been shown to be accurate enough.
GW spectral truncation criteria have been discussed, 
and computationally efficient tools/techniques have been
introduced for constructing reliable templates.
We stress that the above tools/techniques 
could be readily extended, 
to  higher order PN models with relative ease.

\section*{ACKNOWLEDGEMENTS}

V. Pierro has been a Visiting Scientist at the European Space Research \& Technology Centre
ESTEC-ESA, under a grant from the University of Salerno; 
A.D.A.M. Spallicci, formerly staff at ESTEC-ESA,  has been a Visiting Professor 
at the University of Salerno in 1996. 
Both wish to express their appreciation to the hosting Institutions.

\section*{REFERENCES}

{\nid Abramowitz M., Stegun I.A., 1968, {\em Handbook 
of Mathematical Functions}, Dover, New York.}
\\
{Th. Apostolatos, 1996, Phys. Rev. {\ul D52}, 605, 1996. }
\\
{\nid Barone F. et al., 1988, Astron. Astrophys., \ul{199}, 161.}
\\
{\nid Bertotti B. et al., 1999, Phys. Rev. {\ul D59}, 082001, 1999.}
\\
{\nid Eggleton P.P., 1983, Ap. J. \ul{268}, 368.}
\\
{\nid Hils D. et al., 1992, Ap. J. \ul{360}, 75.}
\\
{\nid K. Martel and E. Poisson,  Phys. Rev. {\ul D60}, 124008, 1999.
\\
{\nid Moreno-Garrido C.  et al., 1994, MNRAS, \ul{266}, 16.}
\\
{\nid Moreno-Garrido C.  et al., 1995, MNRAS, \ul{274}, 115.}
\\
{\nid Peters P.C., 1964, Phys. Rev., \ul{136}, 4B, 1124.}
\\
{\nid Peters P.C., Mathews J., 1963, Phys. Rev., \ul{131}, 435. }
\\
{\nid Pierro V., Pinto I., 1996a, Nuovo Cimento B \ul{111}, 631.}
\\
{\nid Pierro V., Pinto I., 1996b, Nuovo Cimento B \ul{111}, 1517.}
\\
{\nid Pierro V., Pinto I., 1996c, Ap. J., \ul{469}, 272.}
\\
{\nid Poisson E. , 1993, Phys. Rev D, \ul{47}, 1497.}
\\
{\nid Press W.H. et al., 1992, {\em Numerical Recipes}, Cambridge Univ.
Press.}
\\
{\nid Taylor J.H. et al., 1993, 
Ap. J. Suppl. Ser., \ul{88}, 529.}
\\
{\nid Wahlquist H., 1987, Gen. Rel. Grav., \ul{19}, 1101.}
\\
{\nid Watson G.N., 1966, {\em A Treatise on the Theory of Bessel 
Functions}, Cambridge Un. Press.}
\\
{\nid  Schott G.A., {\em Electromagnetic Radiation}, Cambridge, 1912.}
\\
{\nid  Prudnikov A.P. et al., {\em Integrals and Series}, Gordon and Breach, 1986.}
\[
~
\]
\[
~
\]
\section*{APPENDIX  A:  RELEVANT TO EQ.S (10) TO (16).}
\[
~
\]
In the weak-field slow-motion approximation,
the cartesian far-field harmonic-gauge metric tensor deviation
components in $(8)$, $(9)$ are simply related 
to the source quadrupole tensor $I_{ij}$ through:
$$
h_{xy}=\frac{2 G}{c^4 r} \frac{d^2 I_{xy}}{dt^2},~~~
h_{xx}=\frac{2 G}{c^4 r} \frac{d^2 I_{xx}}{dt^2},~~~
h_{yy}=\frac{2 G}{c^4 r} \frac{d^2 I_{yy}}{dt^2},
\eqno{(A1)}
$$
where:
$$
I_{xx}=\mu  \rho^2  \cos^2(\phi),~~~
I_{yy}=\mu \rho^2  \sin^2(\phi),~~~
I_{xy}=\mu \rho^2 \cos(\phi)\sin(\phi),
\eqno{(A2)}
$$
$\rho$ being the companion star separation,  $e$ the eccentricity,
$\phi$ the true anomaly and $\mu$ the reduced mass.\\

The relevant terms of the (reduced) quadrupole moment
can be conveneintly rewritten:
$$
I_{xx}=\mu a^2~\xi^2,~~~
I_{yy}=\mu a^2~\eta^2,~~~
I_{xy}=\mu a^2~\xi\eta,
\eqno{(A3)}
$$
where $a$ is the orbit semimajor axis,
$$
\xi=\left(
\frac{\rho~\cos\phi}{a}
\right),~~~
\eta=\left(
\frac{\rho~\sin\phi}{a}
\right).
\eqno{(A4)}
$$
Then, using the well known Keplerian equations (see, e.g.,
Watson, 1976, ch. XVII)
$$
\frac{\rho~\cos\phi}{a}=\cos E-e,~~~
\frac{\rho~\sin\phi}{a}=(1-e^2)^{1/2}~\sin E,
\eqno{(A5)}
$$
where $E$ is the eccentric anomaly, 
and the relation between the latter and the mean
anomaly $M$,
$$
M=\frac{2\pi t}{T}=E-e\sin E,
\eqno{(A6)}
$$
one can expand $\xi^2$ ,$\eta^2$ and $\xi\eta$ into
Fourier series of argument $M$, taking properly 
into account their parities, viz.:
$$
\xi^2=
\frac{\gamma_0}{2}+
\sum_{n=1}^{\infty}
\gamma_n \cos(nM),
\eqno{(A7)}
$$
$$
\eta^2=
\frac{\delta_0}{2}+
\sum_{n=1}^{\infty}
\delta_n \cos(nM),
\eqno{(A8)}
$$
$$
\xi\eta=
\sum_{n=1}^{\infty}
\eta_n \sin(nM).
\eqno{(A9)}
$$
The relevant Fourier coefficients are readily found.
Hence, using $(A5)$ and $(A6)$:
$$
\gamma_n=
\frac{2}{n\pi}
\int_{0}^{\pi}
\sin[n(E-e\sin E)]
(\sin 2E-2e\sin E)
~dE,
\eqno{(A10)}
$$
$$
\delta_n=
-\frac{2}{n\pi}
(1-e^2)
\int_{0}^{\pi}
\sin[n(E-e\sin E)]
\sin 2E~dE,
\eqno{(A11)}
$$
$$
\eta_n=
\frac{2}{n\pi}
(1-e^2)^{1/2}
\int_{0}^{\pi}
\cos[n(E-e\sin E)]
(cos 2E-e \cos E)~
dE.
\eqno{(A12)}
$$
Upon repeated use of  trivial trigonometric
identities, and in view of the integral definition of
the Bessel function of the $1$st kind,
$$
J_{\nu}(\alpha)=
\frac{1}{\pi}
\int_{0}^{\pi}
cos[\nu  x- \alpha \sin x] dx,
\eqno{(A13)}
$$
the Fourier coefficients $(A10)$ to $(A12)$ can be written:
$$
\gamma_n=
\frac{1}{n}
\left[J_{n-2}(ne)-J_{n+2}(ne)\right]
-\frac{2e}{n}
\left[J_{n-1}(ne)-J_{n+1}(ne)\right],
\eqno{(A14)}
$$
$$
\delta_n=
-\frac{1}{n}
(1-e^2)
\left[J_{n-2}(ne)-J_{n+2}(ne)\right],
\eqno{(A15)}
$$
$$
\eta_n=
\frac{1}{n}
(1-e^2)^{1/2}
\left[
J_{n-2}(ne)+J_{n+2}(ne)-2J_n(ne)
\right].
\eqno{(A16)}
$$
Using $(A14)$ to $(A16)$ and $(A7)$ to $(A9)$  in $(A1)$ to $(A3)$ 
gives equations $(10)$ to $(16)$.
\[
~
\]
\[
~
\]
\section*{APPENDIX B: RELEVANT TO EQUATIONS $(22)-(26)$}

In order to establish eq.s $(22)$ to $(26)$ one may repeatedly use the recurrency
formula:

$$
J_{n\pm1}(z) =\frac{n}{z}J_{n}(z) \pm J_{n}^{\prime }(z),
\eqno{(B1)}
$$
so as to reduce the sought series to combinations of the following
(generalized) Kapteyn's expansions of the second kind:

$$
\sum_{n=1}^{\infty}n^{2}[J_{n}^{^{\prime }}(ne)]^{2},
\eqno{(B2)}
$$
$$
\sum_{n=1}^{\infty }[J_{n}^{^{\prime }}(ne)]^{2},
\eqno{(B3)}
$$
$$
\sum_{n=1}^{\infty }\frac{[J_{n}^{^{\prime }}(ne)]^{2}}{n^{2}},
\eqno{(B4)}
$$
$$
\sum_{n=1}^{\infty }n^{2}J_{n}^{2}(ne),
\eqno{(B5)}
$$
$$
\sum_{n=1}^{\infty }J_{n}^{2}(ne),
\eqno{(B6)}
$$
$$
\sum_{n=1}^{\infty }nJ_{n}(ne)J_{n}^{^{\prime }}(ne). 
\eqno{(B7)}
$$
These latter can be summed as follows.
From the Fourier analysis of Kepler motion, the following
equations are readily established (see, e.g., Watson, 1966),
ch. 17.2 :

$$
\cos E=-\frac{e}{2}+2\sum_{n=1}^{\infty}
\frac{J_{n}^{^{\prime }}(ne)}{n} \cos nM,
\eqno{(B8)}
$$
$$
\sin E=\frac{2}{e}~\sum_{n=1}^{\infty}
\frac{J_{n}(ne)}{n} \sin nM,
\eqno{(B9)}
$$
$$
\frac{dE}{dM}=(1-e\cos E)^{-1}.
\eqno{(B10)}
$$
Differentiating eq. $(B8)$ w.r.t. $M$, and using $(B10)$, 
one gets:
$$
\frac{\sin E}{1-e\cos E}=2\sum_{n=1}^{\infty}
J_{n}^{^{\prime }}(ne) \sin nM,
\eqno{(B11)}
$$
$$
\frac{\cos E-e}{(1-e\cos E)^3}=2\sum_{n=1}^{\infty}
n J_{n}^{^{\prime }}(ne) \cos nM.
\eqno{(B12)}
$$
Similarly, from $(B9)$:
$$
\frac{\cos E}{1-e\cos E}=\frac{2}{e} \sum_{n=1}^{\infty}
J_{n}(ne)\cos (nM),
\eqno{(B13)}
$$
$$
\frac{\sin E}{(1-e\cos E) ^{3}}=\frac{2}{e} \sum_{n=1}^{\infty }
n J_{n}(ne)\sin (nM),
\eqno{(B14)}
$$
where $E$ is the eccentric anomaly, $M$ the mean anomaly,
and $e$ the eccentricity. The following procedure can be then
applied to eq.s $(B8)$ and $(B11)$-$(B14)$: {\em i)} squaring;
{\em ii)} taking the average in $M$ over $(0,2\pi)$, using again
eq. $(B10)$; {\em iii)} using the well known (Euler) transformations:
$$
\cos E=(z+z^{-1})/2,~~~\sin E=-i(z-z^{-1})/2,~~~dE=-iz^{-1}dz,
$$
so as to express the sought  series as  contour
integrals on $|z|\!=\!1$ of  rational functions of $z$, which are trivially
computed in terms of residues. 

As an example, applying the above procedure to eq. $(B13)$,
one gets:

$$
\frac{4}{e^2}
\sum_{n=1}^{\infty}
J_n^2(ne)=
\frac{1}{2\pi}
\int_{0}^{2\pi}
\frac{\cos^2 E}{1-e\cos E}~dE=
\frac{1}{2\pi i}
\oint_{|z|=1}
\frac{(1+z^2)^2}{z^2[4z-2e(1+z^2)]}~dz=
$$
$$
=\sum_{|z_i|<1}~Res
\left[
\frac{(1+z^2)^2}{z^2[4z-2e(1+z^2)]}
\right]_{z=z_i}.
\eqno{(B15)}
$$
The integrand function on the r.h.s. of $(B15)$ has a double pole
at $z=0$ and two simple ones at $z=(2e)^{-1}[1\mp(1-e^2)^{1/2}]$.
Only two poles above fall within $|z|<1$, and $(B15)$ gives:
$$
\sum_{n=1}^{\infty}
J_n^2(ne)=
\frac{1}{2}\left[
(1-e^2)^{-1/2}-1
\right]
\eqno{(B16)}
$$
in agreement with Watson, ch. 17.6, eq. (2).  Similarly, starting from
$(B8)$, $(B11)$, $(B13)$ and $(B14)$ one gets, respectively \footnote{Note that
equation (3)  in Watson ch. 17.6,  is in error, as seen by comparison
with $(B20)$, and by direct numerical check. For this (erroneous) result
Watson quotes (Schott, 1912). The same error appears in 
(Prudnikov et al., 1986, sect. 5.7.31).}:
$$
\sum_{n=1}^{\infty }\frac{[J_{n}^{^{\prime }}(ne)]^{2}}{n^{2}} =\frac{1}{2}%
\left( 1-\frac{e^{2}}{4}\right) , 
\eqno{(B17)}
$$
$$
\sum_{n=1}^{\infty }[J_{n}^{^{\prime }}(ne)]^{2} =\frac{1}{2e^{2}}\left[
1-(1-e^{2})^{1/2}\right] , 
\eqno{(B18)}
$$
$$
\sum_{n=1}^{\infty }n^{2}[J_{n}^{^{\prime }}(ne)]^{2} =\frac{4+3e^{2}}{%
8\left( 1-e^{2}\right) ^{5/2}},
\eqno{(B19)}
$$
$$
\sum_{n=1}^{\infty }n^{2}J^{2}_{n}(ne)=\frac{e^{2}(4+e^{2})}{16\left(
1-e^{2}\right) ^{7/2}}.
\eqno{(B20)}
$$

The series  $(B7)$ can be summed by differentiating both sides of 
eq. $(B16)$ w.r.t. $e$. Hence:
$$
\sum_{n=1}^{\infty }nJ_{n}(ne)J_{n}^{^{\prime }}(ne)=\frac{e}{4\left(
1-e^{2}\right) ^{3/2}}.
\eqno{(B21)}
$$
\[
~
\]
\[
~
\]
\section*{APPENDIX C: GENERALIZED CARLINI-MEISSEL EXPANSIONS}
\[
~
\]
To obtain the generalized Carlini Meissel expansion for
$J_{n\pm k}(ne)$ we start from Bessel equation
for $J_{n\pm k}(ne)$:

$$
\frac{d^2J_{n\pm k}(ne)}{de^2}+
\frac{1}{e}\frac{dJ_{n\pm k}(ne)}{de}+
\left[n^2-\frac{(n\pm k)^2}{e^2}\right]J_{n\pm k}(ne) = 0,
\eqno{(C1)}
$$
and let\footnote{This formula is
suggested by the well-known McLaurin
expansions of Bessel functions.}:

$$
J_n(ne)~=~
J_{n\pm k}(ne) = \frac{n^{(n \pm k)}}{(n \pm k )!}
\exp \left[\int_{0}^{e} u_{n \pm k}(z)\right].
\eqno{(C2)}
$$
On letting eq.s $(C2)$ into $(C1)$, we get:

$$
\dot{u}_{n\pm k}+
u^{2}_{n\pm k}+
e^{-1}u_{n\pm k}+
n^{2}-\frac{(n\pm k)^{2}}{e^{2}}=0~,
\eqno{(C3)}
$$
then, following Carlini and Meissel, we assume that 
the following {\em asymptotic} representation
for $u_{n\pm k},~(k=0,1,2)$ holds:

$$
u_{n\pm k}(z)\approx
\frac{u_{n\pm k}^{-}(z)}{n}+
u_{n\pm k}^{0}(z)+
n~u_{n\pm k}^{+}(z)~.
\eqno{(C4)}
$$
Substituting $(C4)$ into $(C3)$, and equating
like powers of $n$ (as required by consistency), we get:

$$
u_{n\pm k}^{+}=\frac{\sqrt{1-z^{2}}}{z},
\eqno{(C5)}
$$
$$
u_{n\pm k}^{0}=\frac
{
\pm 2k - 
e~u_{n\pm k}^{+} -
e^2~\dot{u}_{n\pm k}^{+}
}
{2e^2~u_{n\pm k}^{+}},
\eqno{(C6)}
$$

$$
u_{n\pm k}^{-}=
\frac
{
k^2-
e~u_{n\pm k}^{0} -
e^2~\left[
         \dot{u}_{n\pm k}^{0}+
         (u^{0}_{n\pm k})^{2}
    \right]
}
{2e^2~u_{n\pm k}^{+}}.
\eqno{(C7)}
$$
Hence:

$$ 
u_{n\pm k}^{0}=
\frac{z}{2(1-z^2)}\pm k~\frac{1}{z~\sqrt{1-z^2}},
\eqno{(C8)}
$$

$$
u_{n\pm k}^{-}=
\frac{-z^3-4z}
     {8(1-z^2)^{5/2}}
\mp k~\frac{z}{(1-z^2)^2}+
k^2~\frac{z^3-z}{2(1-z^2)^{5/2}}.
\eqno{(C9)}
$$
Carrying out the integrations in  $(C2)$, and taking into 
account that   $J_{n\pm k}(0)=\delta_{n\pm k,0}$ 
we get:

$$
\int^{e}~u^{+}_{n\pm k}(z)~dz=\sqrt{1-e^2}
+log\left[\frac{e}{1+\sqrt{1-e^2}}\right]+C^{+}_{n\pm k},
\eqno{(C10)}
$$

$$
\int^{e}~u^{0}_{n\pm k}(z)~dz=-\frac{1}{4}~log~(1-e^2)
\pm k~log\left[\frac{e}{1+\sqrt{1-e^2}}\right]+C^{0}_{n\pm k},
\eqno{(C11)}
$$

$$
\int^{e}~u^{-}_{n\pm k}(z)~dz=\frac{-3e^2-2}{24(1-e^2)^{3/2}}
\mp k~\frac{1}{2(1-e^2)}
+k^2~\frac{e^2-1}{2(1-e^2)^{3/2}}+C^{-}_{n\pm k}.
\eqno{(C12)}
$$

Plugging the last three eq.s into eq. $(C2)$ we obtain:

\[
J_{n\pm k}(ne)=\frac{(ne)^{n\pm k}}{(n\pm k)!}
(1+\sqrt{1-e^2})^{-(n\pm k)}~(1-e^2)^{-1/4}\cdot
~exp\left\{
n\sqrt{1-e^2}
+n^{-1}
\left[
\frac{-3e^2-2}{24(1-e^2)^{3/2}}+
\right.
\right.
\]
$$
\left.
\left.
\mp k~\frac{1}{2(1-e^2)}~-~
k^2~\frac{1-e^2}{2(1-e^2)^{3/2}}
\right]~
+n~C_{n\pm k}^{+}+C_{n\pm k}^{0}+n^{-1}C^{-}_{n\pm k}
\right\}.
\eqno{(C13)}
$$
The unknown integration constants can be found by enforcing
the following obvious asymptotic equality, valid for all
$n$:

$$
J_{n\pm k}~(e \longrightarrow 0) \sim 
\frac{(ne/2)^{(n\pm k)}}{(n\pm k)!}.
\eqno{(C14)}
$$
Hence:

$$
\left\{
\begin{array}{l}
1+C^{+}_{n\pm k}=0,\\
~\\
C^{0}_{n\pm k}=0,\\
~\\
-1/12\mp k/2-k^2/2+C^{-}_{n\pm k}=0.
\end{array}
\right.
\eqno{(C15)}
$$
Hence, from $(C13)$:

\[
J_{n\pm k}(ne)~\approx~
\frac{
\displaystyle{\left(\frac{ne}{2}\right)}^{n \pm k}}{(n\pm k)!}
\left(\frac{1+\sqrt{1-e^2}}{2}\right)^{-(n\pm k)}~(1-e^2)^{-1/4}\cdot
exp\left\{
n\left[\sqrt{1-e^2}-1\right]+
n^{-1}
\left[
\frac{-3e^2-2}{24(1-e^2)^{3/2}}+
\right.
\right.
\]
$$
\left.
\left.
\mp k~\frac{1}{2(1-e^2)}- 
k^2~\frac{1-e^2}{2(1-e^2)^{3/2}}+
\frac{1}{12}\pm\frac{k}{2}+\frac{k^2}{2}
\right]
\right\}.
\eqno{(C16)}
$$
The r.h.s. of eq. $(C16)$ above will be  henceforth
denoted as $J_{n \pm k}^{CM}(ne)$, and  can be more 
conveniently written as in $(29)$ to $(31)$.

\end{document}